\documentclass[11pt,preprintnumbers,amsmath,amssymb,nofootinbib]{revtex4-1}
\usepackage{exscale}
\usepackage{amsmath}
\usepackage{booktabs}
\usepackage{graphicx}
\usepackage{exscale}
\usepackage{relsize}
\usepackage{graphicx}
\usepackage{dcolumn}
\usepackage{bm}
\usepackage{multirow}
\usepackage{CJK}
\usepackage{diagbox}
\usepackage{subfigure}
\usepackage{slashed}
\usepackage{makecell}
\usepackage{adjustbox}

\begin{document}
\title{Composite nature of $Z_b$ states from data analysis}

\author{Lu Zhang$^{1}$}
\author{Xian-Wei Kang$^{1,2}$}\email{xwkang@bnu.edu.cn}
\author{Xin-Heng Guo$^1$}
\affiliation{{$^{1}$ Key Laboratory of Beam Technology of Ministry of Education, College of Nuclear Science and Technology, Beijing Normal University, Beijing 100875, People's Republic of China\\
$^{2}$ Beijing Radiation Center, Beijing 100875, China
}}

\begin{abstract}
We use a near-threshold parameterization with explicit inclusion of the Castillejo-Dalitz-Dyson poles, which is more general than the effective range expansion, to study the bottomonium-like states $Z_b(10610)$ and $Z_b(10650)$. In terms of the partial-wave amplitude, we fit the event number distribution of $B^{(*)}\bar B^*$ system to the experimental data for these resonances from Belle Collaboration. The data could be described very well in our method, which supports the molecular interpretation. Then the relevant physical quantities are obtained, including the $B^{(*)}\bar{B}^*$  scattering length ($a$), effective range ($r$), and residue squared ($\gamma_s^2$) of the pole in the complex plane. In particular, we find the compositeness can range from about 0.4 up to 1 for the $B\bar B^*$ ($B^*\bar B^*$) component in the resonance $Z_b(10610)$ ($Z_b(10650)$).
\end{abstract}

\maketitle

\section{Introduction}
Searching for exotic meson beyond the conventional quark antiquark ($q\bar q$) configuration will push forward people's understanding of the substructure of matter. As a key topic in hadron physics, both theoretical and experimental physicists have put much effort on the exotic $XYZ$ mesons.  Those charged particles with hidden charm/bottom flavor, excluding the pure $c\bar c$/$b\bar b$, open a new era in hadron physics spectroscopy. In 2011, Belle observed two narrow structures $Z_b(10610)$ and $Z_b(10650)$ in the mass spectra of $\pi^{\pm}\Upsilon(ns), n=1, 2, 3$ and $\pi^{\pm} h_b(mP), m=1, 2$
\cite{Belle:2011aa}. Later they were confirmed in the mass spectra of $B^{(*)}\bar B^*$ \cite{Belle:2012koo,Belle:2015upu}. The charged $Z_b$ should be comprised of at least four valence quarks, not the pure $b\bar b$, which is a prominent feature for assigning them as exotic mesons. Based on the observations of experiment, various theoretical approaches have been used to investigate the properties of these two resonances, such as mesonic molecular states~\cite{Cleven:2011gp,Goerke:2017svb,Wang:2018jlv,Nieves:2011zz,Sun:2011uh,Dias:2014pva,
Mehen:2013mva,Wang:2014gwa}, cusp effect \cite{Dias:2014pva,Bugg:2011jr,Haidenbauer:2015yka}, compact tetraquark~\cite{Ali:2011ug}, quark-gluon hybrid~\cite{Braaten:2014qka}, and hadro-quarkonium~\cite{Dubynskiy:2008mq,Danilkin:2011sh}. For a review, see e.g., Refs.~\cite{Chen:2016qju,Lebed:2016hpi,Esposito:2016noz,Guo:2017jvc}. Different models are typically characterized by assuming different constituents inside the resonances. However, the real world may contain several of those mechanisms, and thus the concept of compositeness will be an important quantity towards the quantitative analysis.

In the present work, we focus on the two charged bottomnium-like resonances, $Z_b(10610)$ and $Z_b(10650)$, reported by the Belle Collaboration through the analysis of the $e^+e^-\rightarrow B\bar{B}\pi^{\pm}, B\bar{B}^*\pi^{\pm}$ and $B^*\bar{B}^*\pi^{\pm}$ processes \cite{Belle:2015upu}.
Their masses and widths are \cite{ParticleDataGroup:2020ssz}
\begin{equation}
\begin{split}
M_{Z_{b}}&=10607.2\pm2.0\,\, \text{MeV},\quad \Gamma_{Z_{b}}=18.4\pm2.4\,\, \text{MeV};\\
M_{Z'_b}&=10652.2\pm1.5\,\, \text{MeV},\quad \Gamma_{Z_{b^{\prime}}}=11.5\pm2.2\,\, \text{MeV}.
\end{split}
\label{eqmass}
\end{equation}
For simplicity, we may use $Z_b$ and $Z'_b$ to denote the resonance $Z_b(10610)$ and $Z_b(10650)$ respectively.
We concern the relation of $Z_b$ states and the $S-$wave $B^{(*)}\bar B^*$ scattering given their quantum numbers \footnote{We also notice that the information of quantum number can be accessed by the strong decay in $^3P_0$ quark model (e.g., Ref.~\cite{Feng:2021igh}).} The transitions $Z_b^{\pm}(10610)\rightarrow [B\bar{B}^*+c.c]^{\pm}$ and $Z_b^{\pm}(10650)\rightarrow [B^*\bar{B}^*]^{\pm}$ are found to be dominant channels in corresponding final states, with branching fraction of $(85.6^{+2.1}_{-2.9})\%$ and $(75^{+4}_{-6})\%$, respectively \cite{ParticleDataGroup:2020ssz}. We will also assume the isospin symmetry as done in the experimental analysis, e.g., $(B\bar{B}^*)^+$ system denotes $B^+\bar{B}^{*0}$ and $\bar{B}^0B^{*+}$.

We notice that there is a common feature for many of these exotic states. They lie nearby the thresholds of pairs of open charm or bottom mesons. On account of this, the effective range expansion (ERE) approach may provide a proper framework. Therefore, the use of ERE is extensive in the study of these states in the vicinity of thresholds~\cite{Bethe:1949yr}. However, the appearance of near-threshold zeros of the partial wave amplitude might severely limit the ERE convergence. These zeroes are known as the Castillejo-Dalitz-Dyson (CDD) poles~\cite{Castillejo:1955ed}. In this case, the ERE approach is meaningless for practical applications since its radius is too small. Such issue about the breakdown of ERE was also discussed in e.g., Ref.~\cite{Baru:2010ww}.

In our study, $Z_b$ states are close to $B^{(*)}\bar B^*$ threshold:  $Z_b(10610)$ sits 3 MeV above $B\bar B^*$ threshold, and $Z_b(10650)$ 2.8 MeV above $B^*\bar B^*$ one.  We then employ the more general parameterization with explicit inclusion of a CDD pole. The unknown parameters, especially the ones related to CDD pole, will be fitted to the experimental data, with a very good fitting quality. We then calculate several quantities, such as the residue of pole, $B^{(*)}\bar B^*$ scattering length and effective range. However, the current data quality is not enough to pin down the CDD pole, and the compositeness coefficient $X$ of the two-meson $B\bar B^*$ ($B^*\bar B^*$) component can range from 0.39 (0.36) up to 1 for resonance $Z_b(10610)$ ($Z_b(10650)$). Note that our emphasis is to provide a compositeness value, i.e., the weight of $B^{(*)}\bar B^*$ in the state $Z_b^{(\prime)}$, which is an important concept for quantifying a molecule, but not to claim how good our fit is.

The article is organized as follows. After this introduction, we introduce the partial-wave amplitude $t(E)$ with inclusion of a CDD pole in Sec.~\ref{sects}. In Sec.~\ref{secni}, we present the formula of event distributions for fitting. In Sec.~\ref{secx}, we show the compositeness formula that we used for a resonance.  We demonstrate the fit results for the mass spectra of $B^{(*)}\bar B^*$ in Sec.~\ref{secfit}.  Some relevant physical quantities are also organized here. A summary is given in Sec.~\ref{secsum}.

\section{Two-body scattering $t$-matrix element}
\label{sects}
In this study, our main purpose is to develop a fitting procedure of event number distribution based on partial-wave amplitude and experimental data. Once the unknown parameters are obtained by fitting, we can access to several physical quantities, and especially the compositeness coefficient. Below we will introduce the two-body scattering amplitude that includes explicitly the CDD poles.

The general expression for a partial wave amplitude was discussed in Ref.~\cite{Oller:1998zr}, making use of the $N/D$ method~\cite{Chew:1960iv}.
In principle, the left-hand cut (l.h.c.) contribution could also be considered \cite{Ananthanarayan:2000ht,Dai:2017tew,Dai:2018fmx,Kuang:2020bnk}. However, the l.h.c. contribution due to the pion exchange for heavier-meson scattering is expected to be smaller, which could be treated perturbatively, as already noticed before \cite{Kang:2016ezb}. For a review, one may refer to Ref.~\cite{Oller:2019opk}. Then considering only the contact interaction should work, as a first approximation at least. In such case of only unitary constraint without crossed-channel effect, the two-body scattering $t-$matrix can be expressed as~\cite{Oller:1998zr,Guo:2016wpy}
\begin{equation}
t(s)=\left[\sum_i\frac{\rho_i^2}{s-M_{i,\text{CDD}}^2}+G(s)\right]^{-1}.
\label{ts}
\end{equation}
Every term $\rho_i^2/(s-M_{i,\text{CDD}}^2)$ corresponds to the contribution of one CDD pole, with its residue $\rho_i$ and pole mass $M_{i,\text{CDD}}$.
Since we only consider one resonance each time, one naturally includes only one CDD pole. Then the $S$-wave $B\bar{B}^*$ scattering is given by
\begin{equation}
t(s)=\left[\frac{\rho^2}{s-M_{\text{CDD}}^2}+G(s)\right]^{-1}.
\label{eqts}
\end{equation}
As mentioned, the decays $Z_b\to B\bar B^*$ and $Z_b'\to B^*\bar B^*$ are the dominant channels. We will consider the single-channel scattering case. As a matter of fact, a CDD pole corresponds to a zero of $t(s)$ at $s=M_{\text{CDD}}^2$.  In Ref.~\cite{Chew:1961cer}, the CDD pole is linked to the possibility of unstable elementary particle with the quantum number of the scattering channel. More specifically, if CDD pole lies very far from the threshold, a large compositeness of the meson pairs in the resonance will be implied \footnote{There is an exception, for e.g., $\rho$ meson, CDD pole is at infinity, however, it is a conventional resonance but not a molecule \cite{Oller:1998zr}. }. Otherwise, a CDD pole in the vicinity of the meson pair threshold mostly implies a large portion of elementary state, resulting in a small $X$ value.

The function $G(s)$ in Eq.~\eqref{ts} is the scalar two-point loop function, or simply unitarity loop function. Its expression can be written as \cite{Guo:2016wpy}:
\begin{equation}
G(s)=\alpha(\mu^2)+\frac{1}{(4\pi)^2}\left(\log\frac{m_2^2}{\mu^2}-\varkappa_+\log\frac{\varkappa_+-1}{\varkappa_+}
-\varkappa_-\log\frac{\varkappa_--1}{\varkappa_-} \right),
\label{eqgs}
\end{equation}
with
\begin{equation}
\begin{split}
\varkappa_{\pm}&=\frac{s+m_1^2-m_2^2}{2s}\pm\frac{k}{\sqrt{s}}\,,\\
k&=\frac{\sqrt{(s-(m_1-m_2)^2)(s-(m_1+m_2)^2)}}{2\sqrt{s}}\,,
\end{split}
\end{equation}
where $k$ is the modulus of the center-of-mass three-momentum for a two-partial system with masses $m_1$ and $m_2$. In the present case, $m_1$ denotes the $B$ meson mass and $m_2$ denotes $B^*$ meson mass for the $Z_b$ resonance, and $m_1=m_2=m_{B^*}$ for $Z'_b$. Threshold is defined by $m_{\text{th}}=m_1+m_2$. The constant $\alpha(\mu^2)$ is a subtraction constant, and $\mu$ is the renormalization scale. Note that the term of $\alpha(\mu^2)-\text{log}\mu^2/(16\pi^2)$ is independent of $\mu$. The last two log functions in Eq.~\eqref{eqgs} were split into four pieces in Ref.~\cite{Oller:2006jw}. They are fully equivalent, as can be verified both by mathematical derivation and numerical results.  Such form of $G$ function will be used in the complex plane, as the analytic continuation of the physical value on the real axis of $s$.

Since the states $Z^{(\prime)}_b$ are very close to the threshold of meson pairs $B^{(*)}\bar B^*$, it is convenient to consider its non-relativistic reduction~\cite{Guo:2016wpy}. In this limit the $S$-wave $B\bar{B}^*$ scattering becomes
\begin{equation}\label{eq:ts}
t(E)=8\pi m_\text{th}\left[\frac{\lambda}{E-M_{\text{CDD}}}+\beta-ik\right]^{-1},
\end{equation}
with
\begin{eqnarray}
\lambda&=&\rho^2\frac{8\pi m_\text{th}}{m_\text{th}+M_\text{CDD}},\nonumber\\
\beta&=&8\pi m_\text{th}\alpha(\mu^2)+\frac{1}{\pi}\left(m_1\log\frac{m_1}{\mu}+m_2\log\frac{m_2}{\mu}\right),
\end{eqnarray}
and $k=\sqrt{2\bar\mu(E-m_\text{th})}$ is the three-momentum with $\bar\mu$ denoting the reduced mass. Equation \eqref{eq:ts} fulfils the unitarity constraint $\text{Im}\, t^{-1}(E)=-k/(8\pi m_\text{th})$ vs the relativistic condition  $\text{Im}\, t^{-1}(E)=-k/(8\pi \sqrt{s})$.
In the second Riemann Sheet (RS), which will be indicated by a superscript $\uppercase\expandafter{\romannumeral2}$, we take the expression for $t(E)$ as
 \begin{equation}
t^{\uppercase\expandafter{\romannumeral2}}(E)=8\pi m_\text{th}\left[\frac{\lambda}{E-M_{\text{CDD}}}+\beta+ik\right]^{-1}.
\label{eqt2s}
\end{equation}
Note that there is a change of sign in front of $k$ comparing with $t(E)$ for first (physical) RS in Eq.~\eqref{eq:ts}.
In this convention $k$ is always calculated such that $\text{Im}\,k>0$.

In terms of the CDD pole, we may also obtain the scattering length $a$ and effective range $r$ defined in the effective range expansion:
\begin{equation}
t(E)=8\pi m_\text{th}\left(-\frac{1}{a}+\frac{1}{2}r k^2 -i k\right)^{-1}.
\end{equation}
Then one finds
\begin{equation}
\begin{aligned}
\frac{1}{a}&=-\frac{\lambda}{m_\text{th}-M_{\text{CDD}}}-\beta,\\
r&=-\frac{\lambda}{\mu (m_\text{th}-M_{\text{CDD}})^2}.
\end{aligned}
\label{eqar}
\end{equation}
For a molecular, the effective range $r$ of two-body scattering is typically at the order of 1 fm, dominated by pion exchanges.
Clearly, when $M_{\text{CDD}}\approx m_{\text{th}}$, i.e., the CDD pole position is extremely close to the two-meson threshold, but $\lambda$ keeps at a finite value, one will find $r\to \infty$. It certainly limits the convergence radius of ERE. Then in this circumstance, the ERE approach completely fails and one is obliged to resort to the expression that we considered above.

\section{Event number distribution}
\label{secni}
In this section we consider the event number distribution of the decay process $\Upsilon(10860)\rightarrow [B^{(*)}\bar{B}^*]^{\pm}\pi^{\mp}$ through the $Z^{(\prime)}_b$ intermediate states, namely, the cascade decay chain $\Upsilon(10860)\rightarrow Z_b^{(\prime)\pm}\pi^{\mp}$ and  $Z_b^{(\prime)\pm}\rightarrow [B^{(*)}\bar{B}^*]^{\pm}$. To use a concise notation, we present only the $B\bar B^*$ process for $Z_b$, while
for $Z^\prime_b$ and $B^*\bar B^*$ it is straightforward.

Denote the pole of resonance $Z_b$ as $E_P=M_{Z_b}-i\Gamma_{Z_b}/2$, with $M_{Z_b}$ and $\Gamma_{Z_b}$ being its mass and width, respectively.
As demonstrated in Ref.~\cite{Kang:2016jxw}, the properly normalized differential decay rate for the process of $\Upsilon(10860)\rightarrow [B\bar{B}^*]^{\pm}\pi^{\mp}$ can be written as
\begin{equation}
\frac{d\Gamma_{\Upsilon(10860)\rightarrow B\bar B^*\pi}}{dE}
=\frac{\Gamma_{\Upsilon(10860)\rightarrow Z_b \pi}\Gamma_{Z_b\rightarrow B \bar B^*}}{2\pi|E-M_{Z_b}+i\Gamma_{Z_b}/2|^2},
\label{eqdecay}
\end{equation}
if we are interested in the event distributions with invariant mass around the nominal mass of the $Z_b$, as a narrow resonance.
The variable $E$ is defined as the $B\bar B^*$ invariant mass.

We define a function $d(E)$, which will be exploited to describe the final state interactions of the $B^{(*)}\bar{B}^*$ system as well as the $Z_b$ signal. Near the resonance region, $d(E)$ has the form
\begin{equation}
d(E)\simeq \frac{\gamma_d}{E-E_P},
\label{eqal}
\end{equation}
with its residue $\gamma_d$. Then the differential decay rate Eq.~\eqref{eqdecay} will be
\begin{eqnarray}
\frac{d\Gamma_{\Upsilon(10860)\rightarrow B\bar B^*\pi}}{dE}
&=&\Gamma_{\Upsilon(10860)\rightarrow Z_b \pi}\Gamma_{Z_b\rightarrow B\bar B^*}\frac{|d(E)|^2}{2\pi|\gamma_d|^2}\nonumber \\
&=&\Gamma_{\Upsilon(10860)}\mathcal{B}_{Z_b}\frac{\Gamma_{Z_b}|d(E)|^2}{2\pi|\gamma_d|^2},
\label{eqga}
\end{eqnarray}
where $\mathcal{B}_{Z_b}$ is the product of the branching fractions for the decays $\Upsilon(10860)\rightarrow Z_b \pi$ and $Z_b\rightarrow B\bar{B}^*$:
\begin{equation}
\mathcal{B}_{Z_b}=\frac{\Gamma_{\Upsilon(10860)\rightarrow Z_b\pi}\Gamma_{Z_b\rightarrow B\bar B^*}}{\Gamma_{\Upsilon(10860)}\Gamma_{Z_b}}
=\mathcal{B}(\Upsilon(10860)\rightarrow Z_b\pi)\mathcal{B}(Z_b\rightarrow B\bar B^*).
\end{equation}

In fact, the combination $\Gamma_{Z_b}|d(E)|^2/(2\pi |\gamma_d|^2)$ could provide a standard normalized non-relativistic Breit-Wigner (BW) parametrization,
and we redefine it as
\begin{equation}
\frac{d\hat{M}}{dE}=\frac{\Gamma_{Z_b}|d(E)|^2}{2\pi|\gamma_d|^2}.
\end{equation}
Its normalization integral is given by
\begin{equation}
\mathcal{N}=\int_{-\infty}^{+\infty}dE\frac{d\hat{M}}{dE}.
\end{equation}
By considering the form of $d(E)$ in Eq.~\eqref{eqal}, one will find $\mathcal{N}$ is equal to 1 for a very narrow resonance or bound state,
but not for the case of virtual state or
other situations where the final-state interaction function $|d(E)|$ has a shape that strongly departs from a
non-relativistic BW form.

A specific form of parametrization for $d(E)$ can be introduced by the $B\bar{B}^*$ $S$-wave scattering amplitude $t(E)$.
One has to get rid of the CDD pole (or the zero) of $t(E)$ by dividing it by $(E-M_\text{CDD})/\lambda$ and ends with the new function $d(E)$ as
\begin{equation}
d(E)=\frac{1}{1+\frac{E-M_\text{CDD}}{\lambda}(\beta-ik)}.
\label{eqde}
\end{equation}
The extra factor of $8\pi m_\text{th}$ is also removed, which can be accounted for by the overall normalization constant in the fit function.
It is the $d(E)$ function in Eq.~\eqref{eqde} that is used to describe the final state interaction in a general sense. The shape of $|d(E)|$ is certainly beyond the pure BW one, as an extension. The residue $\gamma_d$ can be calculated by
\begin{eqnarray}
\gamma_d=\frac{\xi}{1+\xi \eta},
\end{eqnarray}
with
\begin{eqnarray}
\xi&=&-i\frac{k_P(\beta-i k_P)}{\mu},\\ \nonumber
\eta&=&\frac{\beta-ik_P}{\lambda},
\end{eqnarray}
and $k_P$ is the three-momentum calculated at the pole position $k_P=\sqrt{2\mu(E_P-m_\text{th})}$ in the RS II for a resonance, that is, the phase of the radicand
is between $[2\pi, 4\pi)$.

For the fitting purpose, we consider data on event distributions for $B\bar{B}^*$ from $e^+e^-\rightarrow B\bar{B}^*\pi$ decays measured by the Belle collaboration~\cite{Belle:2015upu}.
Deduced from Eq.~\eqref{eqga}, one has the differential distribution of signal event number
\begin{equation}
\frac{dN_\text{sig}}{dE}=N_{\Upsilon(10860)}\times\mathcal{B}_{Z_b}\times\frac{\Gamma_{Z_b}|d(E)|^2}{2\pi|\gamma_d|^2},
\end{equation}
with $E$ denoting the invariant mass of $B\bar B^*$.

Here we show the estimate of the $\Upsilon(10860)$ event number accumulated by the Belle detector.
To that end, we exploit the relation between the cross section $\sigma(e^+e^-\rightarrow \Upsilon(10860))$ and the di-electron decay width $\Gamma(\Upsilon(10860)\rightarrow e^+e^-)$: \footnote{A detailed derivation and discussion can be found in Ref.~\cite{Peskin:1995ev}.}
\begin{equation}
\sigma(e^+e^-\rightarrow \Upsilon(10860))=4\pi^2 \cdot\frac{3\Gamma(\Upsilon(10860)\rightarrow e^+e^-)}{M_{\Upsilon(10860)}}\cdot\delta(E_\text{cm}^2-M^2),
\end{equation}
with $M=10.885$ \text{GeV} being the mass of $\Upsilon(10860)$, and $E_\text{cm}$ denoting the center of mass energy.
Noticing
\begin{equation}
\delta(E_\text{cm}^2-M_{\Upsilon(10860)}^2)\approx\frac{1}{\pi M_{\Upsilon(10860)}\Gamma_{\Upsilon(10860)}},
\end{equation}
the expression of cross section will be approximated by
\begin{equation}
\sigma(e^+e^-\rightarrow \Upsilon(10860))\approx\frac{12\pi}{M_{\Upsilon(10860)}^2} \mathcal{B}(\Upsilon(10860)\rightarrow e^+e^-),
\end{equation}
with the branching fraction $\mathcal{B}(\Upsilon(10860)\rightarrow e^+e^-)=(8.3\pm2.1)\times 10^{-6}$ \cite{ParticleDataGroup:2020ssz}.
The three-body $\Upsilon(10860)\rightarrow [B^{(*)}\bar{B}^{(*)}]^{\pm}\pi^{\mp}$ decays and the relevant $Z_b^{(\prime)}$ signals are based on the
121.4 fb$^{-1}$ data accumulated by the Belle detector at a center-of-mass energy near the $\Upsilon(10860)$. Combining these pieces together,
we find the total event number of $N_{\Upsilon(10860)}\approx 1.248\times 10^8$.

Following Ref.~\cite{Belle:2015upu}, the background contribution is parameterized as
\begin{equation}
B(E)=b_0e^{-\alpha(E-(m_B+m_{B^*}))}\epsilon(E) F_\text{PHSP}(E),
\end{equation}
where $b_0$ and $\alpha$ are fitting parameters. The experimental reconstruction efficiency is described by $\epsilon(E)\sim\exp[(E-m_0)/\Delta_0](1-E/m_0)^{3/4}$, with $m_0=10.718\pm0.001\text{GeV}/c^2$ the efficiency threshold and $\Delta_0=0.094\pm0.002 \text{GeV}/c^2$. The $F_{\text{PHSP}}(E)$ function is the phase space for the three-body decay $\Upsilon(10860)\rightarrow B\bar{B}^*\pi$ and reads
$F_{\text{PHSP}}(E)\sim |\vec{P}_B||\vec{P}_{\pi}|$, where $|\vec{P}_B|$ is the modulus of the three-momentum of $B$ in center of mass frame of $B\bar B^*$, and $|\vec{P}_\pi|$ is the modulus of the three-momentum of $\pi$ in the rest frame of $\Upsilon(10860)$. Their explicit forms are
\begin{equation}
\begin{aligned}
|\vec P_B|&=\frac{1}{2E}\left[\left(E^2-(m_B+m_{B^*})^2\right)\left(E^2-(m_{B}-m_{B^*})^2\right)\right]^{1/2},\\
|\vec P_{\pi}|&=\frac{1}{2M_{\Upsilon(10860)}}\left[\left(M_{\Upsilon(10860)}^2-(E+m_{\pi})^2\right)\left(M_{\Upsilon(10860)}^2-(E-m_{\pi})^2\right)\right]^{1/2}.
\end{aligned}
\end{equation}
The signal function should be convoluted with the experimental energy resolution $R(E^{\prime},E)$, which
is described by a Gaussian function
\begin{equation}
R(E^{\prime},E)=\frac{1}{\sqrt{2\pi}\sigma}\text{exp}\left(-\frac{(E^{\prime}-E)^2}{2\sigma^2}\right).
\end{equation}
Following the experimental paper, Ref.~\cite{Belle:2015upu}, we take $\sigma=6.0~\text{MeV}/c^2$ \footnote{The efficiency $\epsilon(E)$ and $\sigma=6$ MeV are applied for both $B\bar B^*$ and $B^* \bar B^*$ spectra, for $M_\text{miss}(\pi)$ lying in the range of 10.59 GeV to 10.73 GeV \cite{Belle:2015upu}.}.

The event number as a function of $B\bar{B}^*$ invariant mass will be written as
\begin{equation}
N(E)=N_{\Upsilon(10860)}\epsilon(E)\int_{m_\text{th}}^{M_{\Upsilon(10860)}-m_\pi}[Y_{Z_b}|d(E')|^2+B(E')] F_{\text{PHSP}}(E') R(E^{\prime},E) dE'\,,
\label{eqint}
\end{equation}
with $Y_{Z_b}=\mathcal{B}_{Z_b}\Gamma_{Z_b}/(2\pi|\gamma_d|^2)$, as a normalization constant to be fitted.
Equation \eqref{eqint} constitutes our final formula to be fitted to data of $B\bar{B}^*$ mass spectra for the $\Upsilon(10860)\rightarrow B\bar{B}^*\pi$ from the Belle Collaboration \cite{Belle:2015upu}. The function $|d(E)|$ renders a generalization of the BW function that is used in the experimental analysis \cite{Belle:2012koo,Belle:2015upu}.

\section{Compositeness for a resonance}
\label{secx}
The compositeness relation for a bound state is well defined by Weinberg \cite{Weinberg:1962hj,Weinberg:1965zz}. However, for a resonance case, the compositeness value becomes a complex number, which hinders a transparent physical interpretation. Several developments are proposed \cite{Weinberg:1962hj,Baru:2003qq,Hyodo:2011qc,Aceti:2012dd,Sekihara:2014kya,Agadjanov:2014ana}. In order to quantify the statement on the nature of the $Z^{(\prime)}_b$ as $B^{(*)}\bar B^*$ composite resonance, we apply here the theory developed in Ref.~\cite{Guo:2015daa} that allows a probabilistic interpretation of the compositeness relation for resonances under the condition that $\sqrt{\text{Re}s_P}$ is larger than the lightest threshold among the coupled channels. The pole in the complex $s$-plane is defined as $s_P=E_P^2$.  More specifically, once the resonance lies in an unphysical RS that is connected with the physical RS along an interval of the real-$s$ axis lying above the threshold for the channel $i$, the compositeness formula developed in Ref.~\cite{Guo:2015daa} will be applicable. Following Ref.~\cite{Guo:2015daa}, the partial compositeness coefficient for channel $i$ reads
\begin{equation}
X_i=|\gamma_i|^2\left|\frac{dG_i(s)}{ds}\right|_{s=s_P},
\end{equation}
whose sum is the total compositeness $X=\sum\limits_{i=1}^n X_i\leq 1$.  The difference between 1 and the total compositeness $X$ over the open channels considered is the elementariness $Z=1-X$, which measures the weight of all other components in the resonance.
One also notices that $X$ is independent of the subtraction constant $\alpha$ appearing in $G(s)$, which will drop out in the derivative of $dG_i(s_P)/ds$.

As mentioned before, we consider the single channel scattering $B^{(*)}\bar B^*$, and the states $Z_b^{(\prime)}$ considered here fulfils the condition of the above compositeness. Therefore,
\begin{equation}
\label{eq:Xres}
X=|\gamma_s|^2\left| \frac{d G(s)}{d s}\right|_{s=s_P},
\end{equation}
where $-\gamma_s^2$ is the residue of $t(s)$ at the resonance pole position $s_P$ (following convention in Refs.~\cite{Guo:2015daa,Kang:2016zmv}),
\begin{equation}
t^{\uppercase\expandafter{\romannumeral2}}(s)\longrightarrow \frac{-\gamma_s^2}{s-s_P}.
\end{equation}
Equation \eqref{eq:Xres} differs by the bound state formula $X=-\gamma_s^2 \left.\frac{d G(s)}{d s}\right|_{s=s_P}$ only in the additionally introduced absolute value. However, one should also keep in mind that a bound state appears in the first (physical) sheet and resonance appears in second (unphysical) sheet. The situation of $X=1$ corresponds to a pure bound state or molecular. A similar study on the compositeness analysis was recently done in Refs.~\cite{Wang:2022vga,Gao:2018jhk,Du:2021bgb}.

The residue can be calculated by an integration along a closed contour around the pole:
\begin{equation}
\gamma_d=\frac{1}{2\pi i}\oint d(E)dE
\end{equation}
Besides, it is the $d(E)$ function that appears in our fit formula, and one should notice different kinematic factor between $d(E)$, $t(E)$ and $t(s)$. As a result, we have
\begin{equation}
\label{eq:gammas}
\gamma_s^2=-\frac{1}{2\pi i}\oint d(E)dE\times\frac{(E_P-M_\text{CDD})}{\lambda}\times8\pi E_P\times 2E_P.
\end{equation}
We also notice that the difference between $8\pi|E_P|$ and $8\pi m_\text{th}$ is negligible.

We also need to make an analytical extrapolation of $G(s)$ to the second RS. In order to do this we make use of its continuity property \cite{Oller:1997ti} and discontinuity across the real axis:
\begin{equation}
\label{eq:Gcon}
\begin{aligned}
G^{\uppercase\expandafter{\romannumeral2}}(s+i\epsilon)=G^{\uppercase\expandafter{\romannumeral1}}(s-i\epsilon)&=
G^{\uppercase\expandafter{\romannumeral1}}(s+i\epsilon)-2i\text{Im}G^{\uppercase\expandafter{\romannumeral1}}(s+i\epsilon)\\
&=G^{\uppercase\expandafter{\romannumeral1}}(s+i\epsilon)+\frac{i}{4\pi\sqrt{s}}\frac{\sqrt{(s-(m_1-m_2)^2)(s-(m_1+m_2)^2)}}{2\sqrt{s}}.
\end{aligned}
\end{equation}
Equation \eqref{eq:Gcon} is applied to the case of $\text{Im}\,s > 0$. For $\text{Im}\,s < 0$ case, we use the Schwarz reflection principle $G(s^*)=[G(s)]^*$ and get
\begin{equation}
\label{eq:Gcon2}
\begin{aligned}
G^{\uppercase\expandafter{\romannumeral2}}(s-i\epsilon)=G^{\uppercase\expandafter{\romannumeral1}}(s-i\epsilon)
-\frac{i}{4\pi\sqrt{s}}\frac{\sqrt{(s-(m_1-m_2)^2)(s-(m_1+m_2)^2)}}{2\sqrt{s}}.
\end{aligned}
\end{equation}
In fact, there is also a non-relativistic expansion for $G(s)$ \cite{Wang:2022vga}. The difference between the value of  $\left|dG^{\uppercase\expandafter{\romannumeral2}}/ds\right|_{s=s_P}$ and its leading non-relativistic expansion
is less than 5\%.

\section{Results and discussion}
\label{secfit}
In this section, we provide the fitting strategy, given the data of $B^{(*)}\bar B^*$ mass spectrum from Belle collaboration \cite{Belle:2015upu}. Once the parameters are fixed, the quantities $|\gamma_s|$ and the resulting compositeness $X$ can be calculated. If pertinent, the relevant scattering length and effective range will also be given. The best values of those parameters will be obtained by the routine MINUIT~\cite{minuit}.
The function to be minimized is defined as
\begin{equation}
\chi^2=\sum_{i=1}^{n}\frac{(N_{\text{th},i}-N_{\text{exp},i})^2}{\sigma_i^2},
\end{equation}
where $N_{\text{exp},i}$ is the $i$th experimentally measured event number with uncertainty $\sigma_i$, and $N_{\text{th},i}$ is the event number calculated theoretically given in Eq.~\eqref{eqint}.

There are 6 parameters in total in Eq.~\eqref{eqint}: $b_0,\alpha$ in the background term, $Y_{Z_b}$ as the normalization constant, $\lambda$, $M_\text{CDD}$ and $\beta$ characterizing the signal function $d(E)$. However, as also found in Ref.~\cite{Kang:2016jxw}, there is redundancy between
$\lambda$, $M_\text{CDD}$ and $\beta$. One could impose more conditions to reduce the number of free parameters. We want $t$-matrix to have a resonance pole with
mass and width that are identical to the experimental value given in Eq.~\eqref{eqmass}. In this way, $\lambda$ and $\beta$ can be expressed by $M_\text{CDD}$, and only one parameter $M_\text{CDD}$ remains in $d(E)$ function. More explicitly, the vanishing of the real and imaginary part of $t^{\uppercase\expandafter{\romannumeral2}}(E_P)^{-1}$ leads to the following relation for $\lambda$ and $\beta$,
\begin{equation}
\begin{split}
\lambda&=\frac{1}{2\Gamma_{Z_b}}(\Gamma_{Z_b}^2+4\Delta^2)v\cos u,\quad \beta=\frac{v}{\Gamma_{Z_b}}(-2\Delta\cos u+\Gamma_{Z_b}\sin u),\\
u&\equiv \frac{1}{2}\arg\left(M_{Z_b}-m_\text{th}+i\frac{\Gamma_{Z_b}}{2}\right),\quad v\equiv\left(\mu\sqrt{4(M_{Z_b}-m_\text{th})^2+\Gamma_{Z_b}^2} \right)^{1/2},\quad  \Delta\equiv M_{Z_b}-M_\text{CDD}.
\end{split}
\label{eqdiff}
\end{equation}
Hence, the fitted parameters will be $b_0, \alpha, Y_{Z_b}, M_\text{CDD}$, and only one parameter $M_\text{CDD}$ determines the signal shape $|d(E)|$. Also, the compositess $X$ is solely determined by $M_\text{CDD}$, i.e., there is a one to one correspondence.

\begin{figure}[ht]
\begin{minipage}{1.0\linewidth}
\centerline{\includegraphics[width=0.5\textwidth]{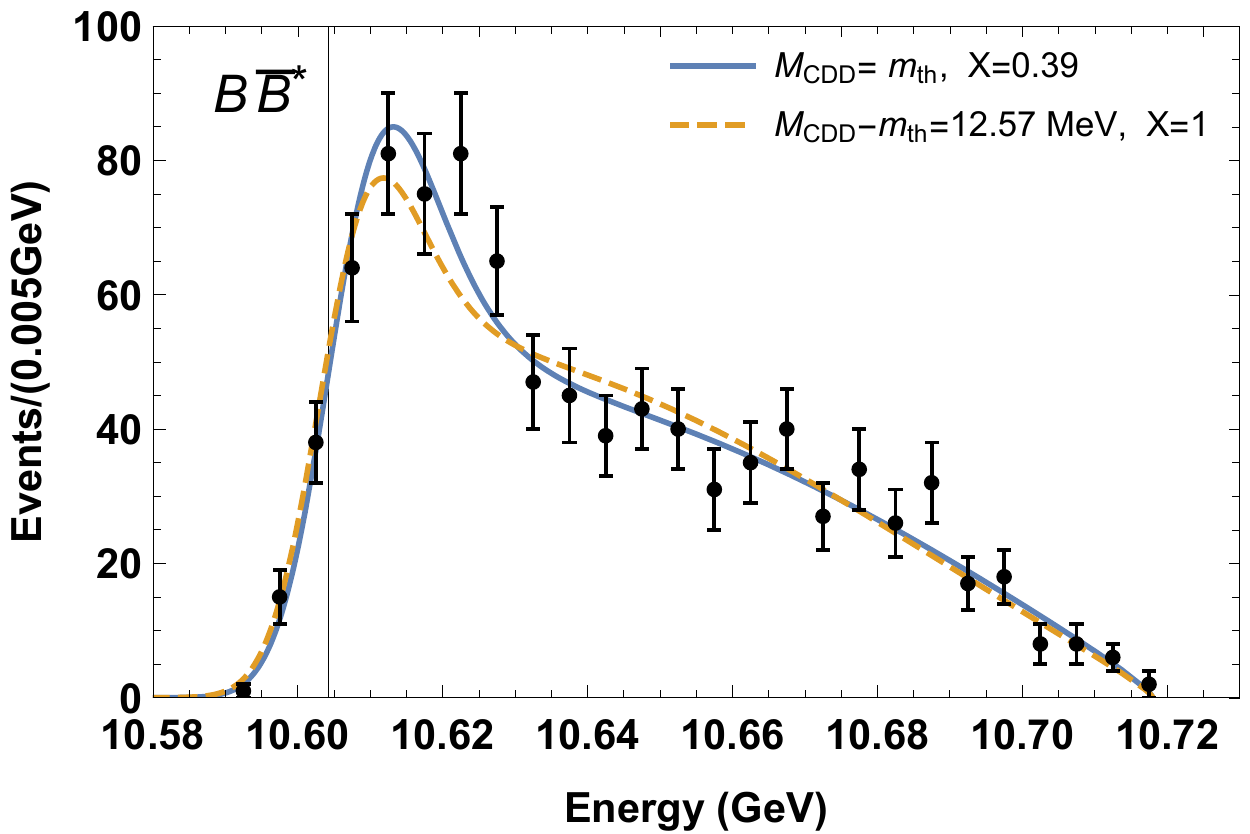}\;\;\;\includegraphics[width=0.5\textwidth]{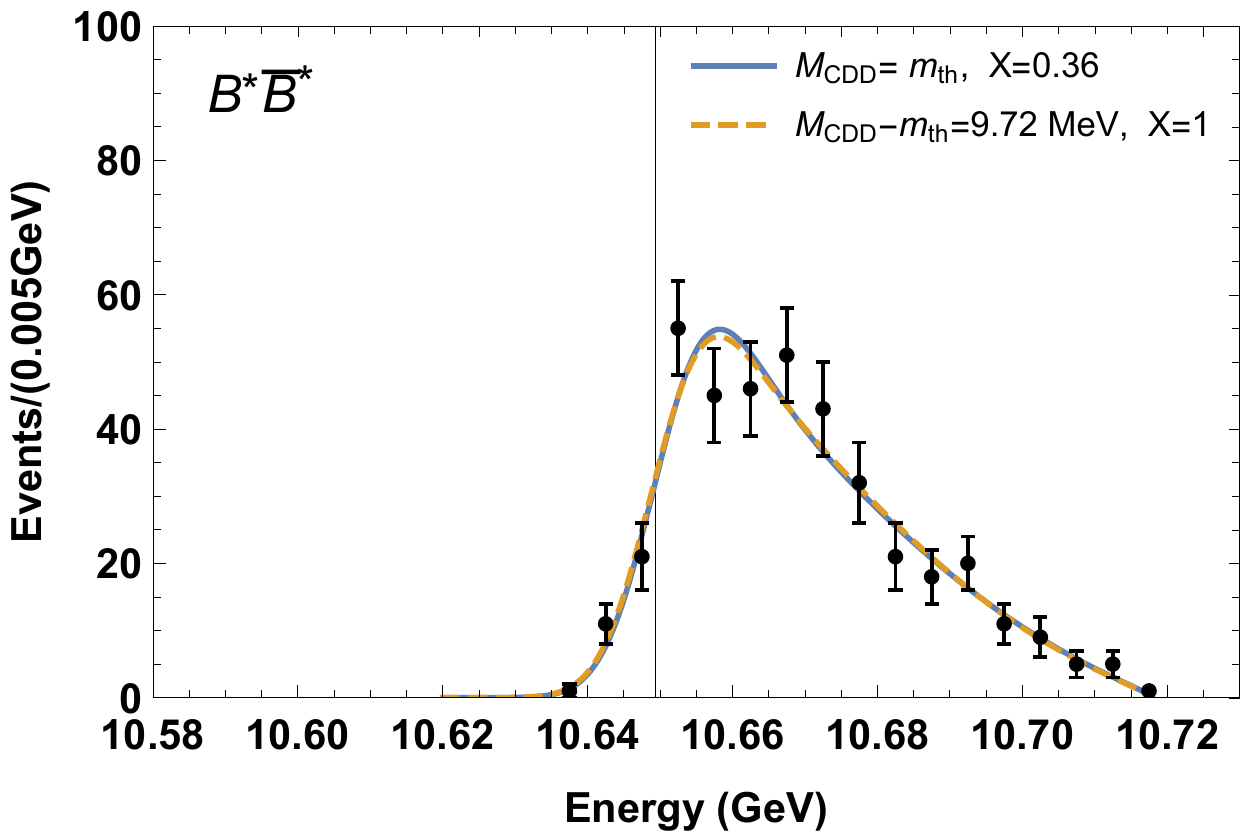}}
\end{minipage}
\caption{Mass spectra of $Z_b(10610)$ and $Z_b(10650)$. The left (right) one is for $B\bar B^*$ ($B^*\bar B^*$) invariant mass spectrum and the relevant $Z_b(10610)$ ($Z_b(10650)$) resonance. Points with error bar are the experimental data from Belle Collaboration \cite{Belle:2015upu}. The vertical lines correspond to the $B^{(*)}\bar{B}^*$ thresholds. In each subfigure, the solid line corresponds to $M_{\text{CDD}}=m_\text{th}$ case, where the value of compositeness $X$ is smallest, 0.39 for $Z_b(10610)$ and 0.36 for $Z_b(10650)$; the dashed line corresponds to $X=1$.  }
\label{figvs}
\end{figure}

We find in our exploration $M_\text{CDD}$ can vary in a wide range, or even across $m_\text{th}$, with all these cases almost describing the data equally well. We then fix $M_\text{CDD}$ to be some special values, either at $m_\text{th}$ in which case the compositeness $X$ achieves its minimum value, or at a value such that $X\approx 1$. Again, we find that they can describe the data very well, with the tunning of normalization constant $Y_{Z_b}$. Then, the values of parameters are organized in Table \ref{tab:para}. The corresponding plots are shown in
Fig.~\ref{figvs}, where the left one is for $B\bar B^*$ invariant mass spectrum and the relevant $Z_b(10610)$ resonance, and the right one is for $B^*\bar B^*$ invariant mass spectrum and the relevant $Z_b(10650)$ resonance. The solid lines correspond to $M_\text{CDD}=m_\text{th}$: $M_\text{CDD}=m_B+m_{B^*}$ for $Z_b(10610)$ and $M_\text{CDD}=2m_{B^*}$ for $Z_b(10650)$. At the point of $M_\text{CDD}=m_\text{th}$, the compositeness has smallest value: $X=0.39$ for $B\bar B^*$ component in $Z_b(10610)$ and $X=0.36$ for $B^*\bar B^*$ component in $Z_b(10650)$. The dashed lines correspond to $X=1$, from which the value of $M_\text{CDD}$ is fixed: $M_\text{CDD}=12.57\,\text{MeV}+m_B+m_{B^*}$ for $Z_b(10610)$ and $M_\text{CDD}=9.72\,\text{MeV}+2m_{B^*}$ for  $Z_b(10650)$. Clearly, both the solid and dashed line match the data well. Especially for $Z_b(10650)$ in the right panel, the two lines are almost indistinguishable. The compositeness coefficient $X$ can roughly range from 0.4 to 1 for both $Z_b$ states by analysing the current experimental data. The value of $1-X$ measures other non-$B^{(*)}\bar B^*$ components, as more elementary degrees of freedom of quantum chromodynamics (QCD), like $b\bar b+\text{gluon}$, compact tetraquark, or their superposition, etc. In Ref.~\cite{Cui:2011fj}, the authors also found that both the interpretations of a molecular state and a tetraquark state are possible by QCD sum rule.

Here we make a comment. The BW shape does describe the data on the invariant mass spectrum well, as done by experimental colleagues. We have checked that the $\chi^2/\text{dof}$ value is 0.81 and 0.94 for $Z_b(10610)$ and $Z_b(10650)$, respectively, which are comparable to the values shown in
Table \ref{tab:para}. In both cases of BW and $|d(E)|^2$ fit, there are only three parameters --- $b_0, \alpha$ in background and $Y_{Z_b}$ as an overall constant. They all describe the mass-spectrum data with the same accuracy basically. However, we are interested in the compositeness value that can be rendered in a model through a careful data analysis, but not claiming that we achieve a good fit.

\begin{table}[!htbp]
\centering
\caption{Fit results for the resonances $Z_b(10610)$ and $Z_b(10650)$. The fitted parameters $b_0, \alpha, Y_{Z_b}$ are given, while $M_\text{CDD}$ are fixed to be some specific values. The quantity of $\chi^2$ per degree of freedom are listed in the last column.}
\renewcommand\arraystretch{1.2}
\begin{tabular}{cp{2cm}<{\centering} p{2cm}<{\centering}p{2.8cm}<{\centering}p{3.2cm}<{\centering}p{1.5cm}<{\centering}p{1.8cm}<{\centering}}
\hline\hline
\;&$b_0$ & $\alpha$\,[GeV$^{-1}$] & $Y_{Z_b}\,[10^{-3}\text{GeV}^{-1}]$ & $M_\text{CDD}-m_\text{th}$\,[MeV]&$\chi^2/\text{dof}$ \\
\hline
\multirow{2}*{$Z_b(10610)$}&$4.87\times10^{-4}$&$10.09$&$3.27$&$0.00$&0.78\\
&$6.32\times10^{-4}$&$13.58$&$0.34$&$12.57$&1.13\\
\hline
\multirow{2}*{$Z_b(10650)$}&$1.01\times10^{-3}$&$33.05$&$1.86$&$0.00$&0.86\\
&$1.11\times10^{-3}$&$35.01$&$0.24$&$9.72$&0.83\\
\hline\hline
\end{tabular}
\label{tab:para}
\end{table}

In Ref.~\cite{Kang:2016ezb}, we have expressed the compositeness $X$ from the scattering parameters, $a$ and $r$, or from the pole mass and width directly. Considering the branching fraction values, we have $X=0.66\pm0.11$ for $Z_b$ and $0.51\pm0.10$ for $Z^\prime_b$ \cite{Kang:2016ezb}.
The values are obviously in the aforementioned range of $0.4\sim 1$. We should stress that the current study of fitting exploits a range of data points of mass spectrum, rather than just a pole position as done in Ref.~\cite{Kang:2016ezb}. That is, more information is considered in our fit,
and the result in Ref.~\cite{Kang:2016ezb} can be regarded as a special case discussed at present.

Based on the results in Table \ref{tab:para}, we will calculate several quantities resulting from them, including $\gamma_s^2$ (the residue squared of $t(s)$ in the $s$ plane, calculated via Eq.~\eqref{eq:gammas}), and the parameters of effective range expansion --- scattering length $a$ and effective range $r$ via Eq.~\eqref{eqar}. Those results are listed in Table \ref{tab:cal}. The values of compositeness $X$ are also repeated there. When the value of $M_\text{CDD}$ coincides with threshold of meson pairs, $a$ and $r$ will encounter infinity and ERE will completely fail. In such case, we do not show the values of $a$ and $r$ in the table.

\begin{table}[!htbp]
\centering
\renewcommand\arraystretch{1.2}
\caption{Several physical quantities are calculated for resonances $Z_b(10610)$ and $Z_b(10650)$, based on the fitted parameters of $b_0, \alpha, Y_{Z_b}$ and given value of $M_\text{CDD}$ in Table \ref{tab:para}. From left to right, the values of $\lambda$, $\beta$, $\gamma_s^2$ (residue squared), $X$ (compositeness), $a$ (scattering length), and $r$ (effective range) are given. When $M_\text{CDD}=m_\text{th}$, the values of $a$ and $r$ do not exist, which is indicated by the symbol ``$-$''.}
\begin{tabular}{cp{2.2cm}<{\centering}p{1.5cm}<{\centering}p{3cm}<{\centering}p{1.5cm}<{\centering}p{1.5cm}<{\centering}p{1.5cm}<{\centering}}
\hline\hline
 &$\lambda\,[\text{MeV}^2]$  &$\beta$\,[MeV]&$\gamma_s^2$\,[GeV$^2$] &$X$&$a$\,[fm]  &$r$\,[fm] \\[1pt]
\hline
\multirow{2}*{$Z_b(10610)$} &$1.87\times10^{3}$  &$73.26$  &$-133.06-i122.49$ &0.39&$-$ &$-$\\[1pt]
&$3.51\times10^{3}$  &$323.73$  &$-235.10+i404.65$ &1.00&$-4.44$ &$-1.65$\\[1pt]
\hline
\multirow{2}*{$Z_b(10650)$} &$1.11\times10^{3}$ &$21.65$  &$-80.69-i110.91$&0.36 &$-$&$-$\\[1pt]
&$2.20\times10^{3}$ &$286.14$  &$-69.37+i380.10$&1.00 &$-3.31$&$-1.72$\\[1pt]
\hline\hline
\end{tabular}
\label{tab:cal}
\end{table}

Finally, we want to remark that the data and resonance mass and width are crucial inputs for our current study of resonance composition. The future improvement of precision of experimental data will certainly constraint better $M_\text{CDD}$ and further $X$, e.g., in the $Z_b(10610)$ case. Besides, some other quantities beyond the mass spectra in $Z^{(\prime)}_b$'s production, e.g., their decay properties could be extremely helpful to pin down their inner structure. As we have experienced in Ref.~\cite{Zhang:2020dla,Haidenbauer:2014kja}, the polarization and forward-backward asymmetry are sensitive to test various model predictions.

\section{Summary}
\label{secsum}
We proposed a general method to take into account the two-body final state interaction near their mass threshold. This method includes the CDD pole
explicitly and thus its applicability range is beyond the traditional effective range expansion (ERE) method. ERE fails when the CDD poles are present very near the threshold.  We then apply this formalism to study the inner nature of the resonances $Z_b(10610)$ and $Z_b(10650)$, with their pole positions lying very close to $B\bar{B}^*$ and $B^*\bar{B}^*$ threshold, respectively.

The free parameters appearing in our equations can be fixed by fitting to experimental data of $B^{(*)}\bar B^*$ mass spectra. Our fitting solution reproduces very well the event number distribution within the whole allowed range of the $B^{(*)}\bar B^*$ invariant mass, which supports the molecular interpretation of these resonances. We find that the value of $M_\text{CDD}$ can vary in a huge range, and both above and below threshold are possible. The resulting compositeness $X$ for $Z_b(10610)$ is $0.39\sim 1$, and $0.36\sim 1$ for $Z_b(10650)$. The study of the compositeness is an important step forward in the quantitative study of hadron structure. The framework adopted here is general for the resonance states located near thresholds of two hadrons. Other related studies are under way. At last, we stress the importance of measuring more observables beyond $B^{*}\bar B^*$ mass spectra, which will be used together to study the property of the resonances $Z_b(10610)$ and $Z_b(10650)$.

\acknowledgments
 The author XWK is indebted to J.~A.~Oller for fruitful discussion, and also for a careful reading and useful comments. It is valuable to have talks with Hong-Rong Qi on some experimental details. We also thank Qian Wang and Christoph Hanhart for their careful readings. This study is supported by the National Natural Science Foundation of China (NSFC) under Project No. 11805012 and No. 11775024.

\section*{References}
\bibliographystyle{apsrev4-1}
\bibliography{ref5}

\end{document}